\documentclass[letterpaper,conference]{IEEEtran}
\IEEEoverridecommandlockouts

\usepackage[utf8]{inputenc}
\usepackage{cite}
\usepackage{amsmath,amssymb,esint}
\usepackage{graphicx,xcolor}
\usepackage[super]{nth}
\usepackage{pbox,ragged2e}
\usepackage{balance}
\usepackage{placeins}

\usepackage[caption=false,font=footnotesize]{subfig}
\usepackage{textcomp}
\usepackage{gensymb}
\usepackage{dsfont}
\usepackage{bbm}
\usepackage{psfrag}

\newcommand{\f}[2]{\frac{#1}{#2}} 

\DeclareMathOperator{\diag}{diag} 

\newcommand{\eye}{\mathbf{I}}

\newcommand{\mtx}[2]{\left[\begin{array}{#1}#2\end{array}\right]}

\newcommand{\Tr}{^\text{T}}
\renewcommand{\H}{^\text{H}}

\newcommand{\EVSymb}{\mathbb{E}} 
\newcommand{\EV}[1]{\EVSymb\{\hspace{.1mm}#1\hspace{.1mm}\}}

  \newcommand{\Z}{{\bf Z}}              
	
  \renewcommand{\v}{\mathbf{v}}         
  \renewcommand{\i}{\mathbf{i}}         
  
	\newcommand{\re}{\mathrm{Re}}
	\newcommand{\im}{\mathrm{Im}}
	
	\newcommand{\inZ}{^{\text{in}}}
	\newcommand{\outZ}{^{\text{out}}}

  \newcommand{\unit}[1]{\,\mathrm{#1}}	
  \newcommand{\um}{\,\text{\textmu}\mathrm{m}}

  \newcommand{\fDesign}{f_\text{D}}

  \newcommand{\ZOpt}{Z_\text{o}}
  
  \newcommand{\T}[1]{_{\mathrm{T#1}}}
  \newcommand{\R}[1]{_{\mathrm{R#1}}}
  \newcommand{\A}[1]{_{\mathrm{A#1}}}
  \newcommand{\Rs}[1]{_{\text{relay}#1}}
  \newcommand{\N}[1]{_{\mathrm{N}#1}}
  \newcommand{\G}[1]{_{\mathrm{G}#1}}
  \renewcommand{\L}[1]{_{\mathrm{L}#1}}
  
  \newcommand{\e}{\mathbf{e}}
\renewcommand{\o}{\mathbf{o}}

\newcommand{\psGD}[4]{\psfrag{#1}{\hspace{#2}\raisebox{#3}{\scriptsize{#4}}}}

 \newcommand{\radFreq}{\omega}

\newcommand{\nf}{_\text{near}}
\newcommand{\ff}{_\text{far}}

\title{Magneto-Inductive Powering and Uplink of In-Body Microsensors: Feasibility and High-Density Effects}
\author{%
\IEEEauthorblockN{Gregor Dumphart, Bertold Ian Bitachon, and Armin Wittneben} 
\IEEEauthorblockA{Communication Technology Laboratory\\ ETH Zurich, 8092 Z\"urich, Switzerland\\
Email: dumphart@nari.ee.ethz.ch, bertold.bitachon@ief.ee.ethz.ch, wittneben@nari.ee.ethz.ch}}

\begin{document}

\maketitle

\begin{abstract}
This paper studies magnetic induction for wireless powering and the data uplink of microsensors, in particular for future medical in-body applications. We consider an external massive coil array as power source (1\,W) and data sink. For sensor devices at 12\,cm distance from the array, e.g. beneath the human skin, we compute a minimum coil size of 150\,{\textmu}m assuming 50\,nW required chip activation power and operation at 750\,MHz. A 275\,{\textmu}m coil at the sensor allows for 1\,Mbit/s uplink rate. Moreover, we study resonant sensor nodes in dense swarms, a key aspect of envisioned biomedical applications. In particular, we investigate the occurring passive relaying effect and cooperative transmit beamforming in the uplink. We show that the frequency- and location-dependent signal fluctuations in such swarms allow for significant performance gains when utilized with adaptive matching, spectrally-aware signaling and node cooperation. The work is based on a general magneto-inductive MIMO system model, which is introduced first.


\end{abstract}
\section{Introduction}\label{sec:intro}
Wireless magnetic induction has well-known applications in RFID and NFC systems but is also used by biomedical implants \cite{Agarwal2017} 
and sensor networks in harsh media \cite{KisseleffTC2015}. The main advantages are: (i) the magnetic near-field penetrates most materials with little interaction, (ii) resonant multi-turn coil designs allow for strong links over mid-range distances, making them well-suited for efficient power transfer \cite{Agarwal2017}
, and (iii) performance gains can be achieved by placing resonant coils (passive relays) nearby \cite{DumphartICC2017,KisseleffTC2015,Slottke2016}.

Enabling \textit{medical microrobots} is an important objective of biomedical research. They are expected to provide untethered diagnostic sensing, treatment and drug delivery in future medical in-body applications. 
The maximum suitable device size ranges from around $3\unit{cm}$ for gastrointestinal applications down to a few $\um$ in capillary vessels. The wireless aspects of medical microrobots are challenging open problems: wireless powering is crucial because battery capacity scales down with volume and, moreover, conventional radio link designs are inappropriate for the data uplink to a sink outside the human body because the waves radiated by a micro-scale $\lambda/2$-antenna would not penetrate any tissue. \cite{Sitti2015,Nelson2010}

Magnetic induction with its outlined advantages and suitability for miniaturization has been proposed for wireless-powered medical small-scale devices \cite{Slottke2016,Gulbahar2017}. Yet, the feasibility and behavior of small magneto-inductive wireless nodes in medical ad-hoc settings is unclear. This holds especially true for medical microrobot \textit{swarms}, which offer great opportunities in envisioned biomedical applications \cite{Sitti2015,ServantNelson2015} and also to the wireless engineer: physical layer \textit{cooperation} between in-body devices allows for an array gain and spatial diversity in the uplink. Furthermore, dense swarms of strongly-coupled resonant coils would give rise to a \textit{passive relaying} effect. Such strong coupling is however associated with resonant mode splitting, i.e. a complicated frequency-selective channel \cite{DumphartICC2017} which should be accounted for by the signaling scheme. Yet, to the best of our knowledge, no existing communications study properly accounts for this channel. A meaningful study of communication range must furthermore consider radiative propagation even when the coils are electrically small.

In regards to these shortcomings, this paper makes the following contributions. 
We present a concise general system model for magneto-inductive networks applicable to both power transfer and communication with an arbitrary number of transmitters, receivers and passive relays, with any possible arrangement or matching circuitry. We account for near- and far-field propagation, antenna coupling, noise correlation and $f$-selectivity by employing methods from MIMO radio communications with compact arrays \cite{IvrlacTCS2010}. 
Based thereon, we conduct a numerical feasibility study of medical in-body nodes in a setup with a large external device that features a massive coil array and serves as power source and data sink. In particular, we study the minimum sensor-side coil size for feasible wireless powering and the resultant uplink data rates. For the case when a sensor is part of a dense swarm of resonant nodes we show that (and how) the resulting $f$-selective fading-type channel can be utilized for performance gains.

Sec.~\ref{sec:model} presents the narrowband signal and noise model which is extended to broadband by Sec.~\ref{sec:broad} while taking practical matching circuits under consideration. We present the considered biomedical setup in Sec.~\ref{sec:setup} and the associated numerical results in Sec.~\ref{sec:effects}. Then Sec.~\ref{sec:sum} concludes the paper.



%
\section{System Model}\label{sec:model}
This section presents a signal and noise model for multiple-input multiple-output (MIMO) systems of electrically small loop antennas (coils) in any arrangement, which is then used to study wireless-powered medical sensors. The model can however serve for any magneto-inductive evaluation (communication, power transfer, or sensing) and, of course, entails any SISO, SIMO and MISO case. The approach closely follows the multiport circuit theory of communication presented in \cite{IvrlacTCS2010} (we refer to this paper for details on the employed concepts).

\begin{figure}[!ht]\centering

\psGD{gen}{-3mm}{.21mm}{\shortstack[c]{'G'\\transmit\\generators\\$1 \ldots N\T{}$}}
\psGD{chan}{-13.4mm}{.21mm}{\shortstack[c]{'A'\\loop antenna coupling via\\near- and far-field propagation,\\coil equivalent circuits}}
\psGD{lna}{-2.9mm}{.21mm}{\shortstack[c]{'L'\\receiver\\loads\\$1 \ldots N\R{}$}}
\psGD{TD}{-4.0mm}{.21mm}{\shortstack[c]{'T'\\transmit\\matching\\network(s)}}
\psGD{RD}{-4.4mm}{.21mm}{\shortstack[c]{'R'\\receiver\\matching\\network(s)}}

\psGD{Rg}{0mm}{2.4mm}{$R$}
\psGD{vGA}{.2mm}{-.5mm}{$v\G{,1}$}
\psGD{vGE}{.2mm}{-.5mm}{$v\G{,N\T{}}$}
\psGD{pG}{-.5mm}{0mm}{$+$}
\psGD{mG}{-.5mm}{0mm}{$-$}

\psGD{MT}{0.0mm}{0.0mm}{$\Z\T{}$}
\psGD{ZA}{0mm}{-1mm}{$\Z\A{}$}
\psGD{MR}{0.0mm}{0.0mm}{$\Z\R{}$}

\psGD{pt}{0mm}{0mm}{}
\psGD{pr}{0mm}{0mm}{}
\psGD{RtA}{2.0mm}{0.0mm}{} 
\psGD{RtE}{2.0mm}{0.0mm}{} 
\psGD{RrA}{-2.0mm}{0.0mm}{} 
\psGD{RrE}{-2.0mm}{0.0mm}{} 
\psGD{LtA}{-5.3mm}{0.0mm}{Tx coil $1$}
\psGD{LtE}{-6.8mm}{-.3mm}{Tx coil\,$N\T{}$}
\psGD{LrA}{-2.0mm}{0.0mm}{Rx coil $1$}
\psGD{LrE}{-2.0mm}{-.3mm}{Rx coil\,$N\R{}$}

\psGD{pL}{0mm}{.5mm}{$+$}
\psGD{mL}{0mm}{-.5mm}{$-$}
\psGD{vLA}{-.6mm}{-.2mm}{$v\L{,1}$}
\psGD{vLE}{-.6mm}{-.2mm}{$v\L{,N\R{}}$}
\psGD{RL}{-3.0mm}{0mm}{$R$}

\newcommand{\myV}{.3mm}
\psGD{RIN}{-2.1mm}{\myV}{$R\eye_{N\T{}}$}
\psGD{ZTI}{-1.4mm}{\myV}{$\Z\T{}\inZ$}
\psGD{ZTO}{.2mm}{\myV}{$\Z\T{}\outZ$}
\psGD{ZAI}{-1.6mm}{\myV}{$\Z\A{}\inZ$}
\psGD{ZAO}{.2mm}{\myV}{$\Z\A{}\outZ$}
\psGD{ZRI}{-1.6mm}{\myV}{$\Z\R{}\inZ$}
\psGD{ZRO}{.5mm}{\myV}{$\Z\R{}\outZ$}
\psGD{RIM}{-1.7mm}{\myV}{$R\eye_{N\R{}}$}

\psGD{dd}{0.0mm}{-1.0mm}{\normalsize$\bf\vdots$}

\includegraphics[width=\columnwidth,trim=0 6 5 -8]{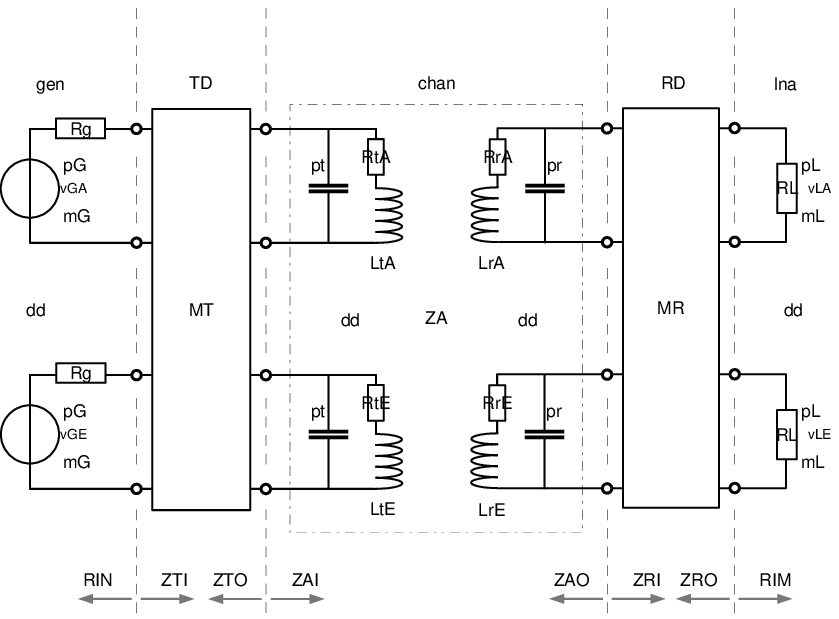}
\caption{Circuit abstraction of $N\T{}$ transmitting and $N\R{}$ receiving electrically small loop antennas (coils) for wireless communication or power transfer. The matching networks can be full multiports or individual two-ports per coil.}
\label{fig:CircSignal}
\end{figure}

Fig.~\ref{fig:CircSignal} shows the considered circuit model with $N\T{}$ transmit (Tx) coils and $N\R{}$ receive (Rx) coils in terms of a circuit description. It is divided into five stages:
transmit generator outputs 'G',
transmitter matching network(s) 'T',
antennas and propagation 'A',
receiver matching network(s) 'R', and
receiver loads 'L'. The latter are either the inputs to low-noise amplifiers (LNAs) of a communication receiver or tank circuits of a power receiver.
Each transmit generator and receiver load has the reference impedance $R = 50\unit{\Omega}$.
The central stages are considered as passive reciprocal multiport networks which have complex symmetric (not Hermitian) impedances matrices
\begin{align}
\Z\T{} &= \mtx{ll}{\Z\T{:G} & \Z\T{:AG}\Tr \hspace{-.35mm} \\ \Z\T{:AG} \hspace{-.3mm} & \Z\T{:A}} \in \mathbb{C}^{2N\T{} \times 2N\T{}} \\
\Z\A{} &= \mtx{ll}{\Z\A{:T} & \Z\A{:TR} \hspace{-.3mm} \\ \Z\A{:RT}   & \Z\A{:R}} \in \mathbb{C}^{(N\T{} + N\R{})\times(N\T{} + N\R{})} \\
\Z\R{} &= \mtx{ll}{\Z\R{:A} & \Z\R{:LA}\Tr  \\ \Z\R{:LA}  & \Z\R{:L}} \in \mathbb{C}^{2N\R{} \times 2N\R{}}.
\end{align}
This abstraction not only comprises the case of co-located coil arrays with multiport matching networks but also distributed nodes with individual two-port networks. 

The goal of this section is a narrowband MIMO model \mbox{${\bf y} = {\bf Hx} + {\bf n}$} at an evaluation frequency $f$ which may differ from the design frequency. Thereby, $\|{\bf x}\|^2$ and $\|{\bf y}\|^2$ shall express physical transmit and receive power.
Guided by \cite{IvrlacTCS2010}, we first state the voltage gain from Tx generators to Rx loads. 

\subsection{Signal Propagation}

For given Tx generator voltages $\v\G{} \in \mathbb{C}^{N\T{}}$ (complex envelope), the Rx load voltages $\v\L{} \in \mathbb{C}^{N\R{}}$ are given by
\begin{align}
\v\L{} = {\bf D} \v\G{} + \v\N{}
\label{eq:VoltageModel}
\end{align}
where $\v\N{} \in \mathbb{C}^{N\R{}}$ are the receiver noise voltages. We express the $N\R{} \times N\T{}$ gain matrix 
as a concatenation of linear signal transfer functions\footnotemark
\newcommand{\myDist}{\hspace{.2mm}}
\begin{align}
{\bf D} = {\bf D}\L{} \myDist {\bf D}\R{} \myDist \Z\A{:RT} \myDist {\bf Y}_\mathrm{T}
\end{align}
which follow from a four-fold application of the Th\'{e}venin theorem for multiport networks. They are given by
\begin{align}
& {\bf D}\L{} = R(R \eye_{N\R{}} + \Z\R{}\outZ)^{-1} \\
& {\bf D}\R{} = \Z\R{:LA} (\Z\R{:A} + \Z\A{}\outZ)^{-1} \\
& {\bf Y}_\mathrm{T} 
= (\Z\A{:T} + \Z\T{:A}   )^{-1} \Z\T{:AG} (\Z\T{}\inZ + R \eye_{N\T{}})^{-1} \label{eq:YTGvB}.
\end{align}
The input and output impedances are stated in the appendix.

\footnotetext{The expression is equivalent to \cite[Eq.~16]{IvrlacTCS2010} apart from the fact that we did not use the unilateral assumption $\Z\A{:TR} \approx \bf 0$ (which is well-justified for weak Tx-Rx coupling, e.g. for radio communications, but not for strong links, e.g. efficient power transfer). Note that verifying the equivalence of some formulas requires (multiple applications of) the Woodbury matrix identity.} 

\subsection{Antenna Self and Mutual Impedances}
\label{sec:coupling}

We consider all coils in terms of their equivalent circuit \cite[Fig.~5.4]{Balanis2016}. The network between all $N\T{}+N\R{}$ coil ports (without matching circuitry) has the impedance matrix
\begin{align}
\Z\A{} = \left(\, \bar\Z^{-1} + j\radFreq \diag{}( C_1^\text{\,self}\!, \ldots , C_{N\T{}+N\R{}}^\text{\,self} ) \right)^{-1}
\label{eq:ZCouplingFinal}
\end{align}
where $\bar\Z$ is the impedance matrix of the same network without the coil self-capacitances $C_n^\text{\,self}$. Its diagonal is given by
\begin{align}
\bar{Z}_{n,n} = R_n^\text{ohm} + R_n^\text{rad} + j\radFreq L_n
\end{align}
where the self-inductance $L_n$ follows from the coil geometry.
$R^\text{ohm}_n$ is determined by the wire length, diameter, material, turn spacing (proximity effect), and frequency (skin effect). The radiation resistance 
$R^\text{rad}_n = \f{1}{3}\,\mu k^3 f \nu_n^2 S_n^2$
with permeability $\mu$, wavenumber $k = 2\pi f/c$ (spatial frequency), turn number $\nu_n$, and coil area $S_n$. Note that both $R^\text{ohm}_n$ and $R^\text{rad}_n$ are frequency-dependent quantities. \cite[Sec.~5.2.3]{Balanis2016}

A pair of electrically small coils of thin wires (curves $\mathcal{C}_m$ and $\mathcal{C}_n$ in 3D space) has mutual impedance \cite[Eq.~5-2]{Balanis2016}
\begin{align}
\bar{Z}_{m,n} = 
\f{j\radFreq \mu}{4\pi} \oint_{\mathcal{C}_m} \oint_{\mathcal{C}_n} \f{e^{-jkd}}{d} d{\bf s}_m \cdot d{\bf s}_n
\label{eq:DoubleIntegralGeneral}
\end{align}
whereby distance $d$ is between a pair of points on $\mathcal{C}_m$ and $\mathcal{C}_n$ and $d{\bf s}_m$ and $d{\bf s}_n$ are infinitesimal line elements. This formula is more general than a magnetoquasistatic description in terms of mutual inductance as it also comprises radiative propagation. This is apparent in the dipole approximation\footnote{This formula is obtained by rearranging the trigonometric field expressions \cite[Eq.~5-18]{Balanis2016} and assuming a spatially constant field across the Rx coil.}
\newcommand{\myD}{\hspace{-.2mm}} 
\begin{align}
& \bar{Z}_{m,n} 
\approx j\radFreq \bar{L} 
\left(\!\left(\f{1}{(kd)^3\!} + \f{j}{(kd)^2} \!\right)\!J\nf + \!\f{J\ff}{2kd} \!\right) \! e^{-jkd}
\label{eq:MutualImpDipole}
\end{align}
with 
$\bar{L} = \f{\mu}{2\pi} \nu_m S_m \nu_n S_n k^3$,
which is accurate when the center-to-center coil separation $d$ is larger than a few times the larger coil diameter.
The unitless $J\nf, J\ff \in [-1,1]$ are given by
\begin{align}
& J\nf = \o_m\Tr \myD \big( \tfrac{3}{2} \e\e\Tr - \tfrac{1}{2} \eye_3 \big)\o_n, \\
& J\ff = \o_m\Tr \myD \big( \eye_3 \myD - \myD \e\e\Tr \big) \o_n.
\end{align}
They account for the coil axis orientations via unit vectors $\o_m$ and $\o_n$ and a center-to-center direction vector $\bf e$.

\subsection{Receiver Noise Statistics}
\label{sec:noise}

To describe the statistics of noise voltage vector $\v\N{}$ in \eqref{eq:VoltageModel} we consider noise signals from various sources and their correlation due to antenna coupling. We employ the well-established assumption of a circularly-symmetric complex Gaussian distribution $\f{1}{\sqrt{R}} \v\N{} \sim \mathcal{CN}({\bf 0},{\bf K})$. The noise covariance matrix (as a power quantity) is composed of \cite{IvrlacTCS2010,HassanWCNC2015}
\begin{align}
{\bf K} = 
\f{1}{R}  \left( {\bf D}\L{} \boldsymbol\Psi {\bf D}\L{}\H + \sigma_\text{iid}^2 \hspace{.2mm}\eye_{N\R{}} \right)
\label{eq:NoiseCovar}
\end{align}
where $\sigma_\text{iid}^2 \hspace{.2mm}\eye_{N\R{}}$ represents noise at the LNA output and later stages; $\sigma_\text{iid}^2$ is the equivalent input voltage variance. The contributions from other sources are characterized by
\begin{align}
& \boldsymbol\Psi = \boldsymbol\Psi_\text{extrinsic} + \boldsymbol\Psi_\text{thermal} + \boldsymbol\Psi_\text{LNA}, \\
& \boldsymbol\Psi_\text{extrinsic} = 4k_\mathrm{B}T_\mathrm{A} W  \cdot {\bf S} \boldsymbol\Phi {\bf S}\H, \\
& \boldsymbol\Psi_\text{thermal} = 4k_\mathrm{B}TW  \cdot \re\{\Z\R{}\outZ \}|_\text{ohmic}, \\
& \boldsymbol\Psi_\text{LNA} = \beta \left( R\N{}^2 \eye_{N\R{}} + \Z\R{}\outZ (\Z\R{}\outZ)\H - 2 R\N{} \re(\rho^* \Z\R{}\outZ) \right).
\label{eq:NoiseCov}
\end{align}
The extrinsic noise statistics are computed from the Boltzmann constant $k_\mathrm{B}$, antenna noise temperature $T_\mathrm{A}$, bandwidth $W$,
matrix
${\bf S} = {\bf D}\R{} {\bf D}_C \diag(R^\text{rad}_1,\ldots,R^\text{rad}_{N\R{}})^\f{1}{2}$
with voltage gain
${\bf D}_C = \eye_{N\R{}} \!- \Z\A{}\outZ \cdot j\radFreq \diag(C^\text{self}_1\!,\ldots,C^\text{self}_{N\R{}})$
past the Rx coil self-capacitances, and spatial correlation matrix $\boldsymbol\Phi$ with unit diagonal. Even for electrically small coils, extrinsic noise can be dominant in unshielded environments due to potentially huge $T_\mathrm{A}$ at low frequencies \cite{RadioNoiseITU2016}.
Thermal noise is due to the ohmic part of $\Z\R{}\outZ$ at (the actual) temperature $T$.
LNA noise is described by the statistics of two equivalent noise sources $v\N{}$ and $i\N{}$ at each LNA input: variance $\beta = \EV{|i\N{}|^2}$, noise resistance $R\N{} = \sqrt{\EV{|v\N{}|^2} / \beta}$ and complex correlation coefficient $\rho = \EV{v\N{} i\N{}^*} / (\beta R\N{})$ with $|\rho| \leq 1$.
\cite[Sec.~II-E]{IvrlacTCS2010}

\subsection{Incorporation of Passive Relays}
\label{sec:relays}

To incorporate the effect of $N\Rs{}$ passive relays into $\Z\A{}$, we write the impedance matrix of all $N\T{}\!+\!N\R{}\!+\!N\Rs{}\!$ coils as
\begin{align}
\mtx{ll}{\Z\A{}|_\text{no\,relays} & \Z_\text{to relays}\Tr \\ \Z_\text{to relays} & \Z_\text{relays}}
\end{align}
and calculate all elements according to Sec.~\ref{sec:coupling}. By terminating the relay ports with a passive load $\Z_\text{term}$ we obtain
\begin{align}
\Z\A{} &= \Z\A{}|_\text{no\,relays} - \Z_\text{to relays}\Tr (\Z_\text{relays} + \Z_\text{term})^{-1} \Z_\text{to relays}
\end{align}
between the $N\T{} + N\R{}$ original coils. A natural choice of load is one resonance capacitor per relay coil. \cite{DumphartICC2017,KisseleffTC2015,Slottke2016} 

\subsection{Matching}
\label{subsec:match}
The active power delivered by the generators to the transmit antennas is maximized by \textit{power matching}, i.e. by enforcing $\Z\T{}\inZ = R\eye_{N\T{}}$ and $\Z\T{}\outZ = (\Z\A{}\inZ)^*$. This is achieved with a matching network that exhibits impedance matrix \cite[Eq.~103]{IvrlacTCS2010}
\renewcommand{\myDist}{\hspace{.6mm}}
\begin{align}
\Z\T{} = j\mtx{cc}{{\bf 0}_{N\T{}} & \pm\sqrt{R}\myDist\re\{\Z\A{}\inZ\}^\f{1}{2} \\ \pm\sqrt{R}\myDist\re\{\Z\A{}\inZ\}^\f{1}{2} & -\im\{\Z\A{}\inZ\}}.
\label{eq:TxPowerMatchMP}
\end{align}
The same principle can be applied at the Rx-side for maximum power transfer efficiency (PTE). 
The SNR at a communication receiver is however maximized with \textit{noise matching} which presents $\Z\R{}\outZ = \ZOpt \eye_{N\R{}}$ with an optimal impedance $\ZOpt = R\N{} ( \sqrt{1 - (\im\,\rho)^2} + j\im\,\rho )$ to the LNA inputs. This is achieved by a matching network with \cite[Sec.~IV.B]{IvrlacTCS2010}
\renewcommand{\myDist}{\hspace{-.2mm}}
\begin{align}
\Z\R{} &= j\mtx{cc}{-\im\{\Z\A{}\outZ\} & \!\!\!\!\! \pm(\re\{\ZOpt\myDist\} \re\{\Z\A{}\outZ\})^\f{1}{2} \!\! \\ 
\!\!\! \pm(\re\{\ZOpt\myDist\} \re\{\Z\A{}\outZ\})^\f{1}{2} \!\!\!\!\! & \im\{\ZOpt\myDist\}\,\eye_{N\R{}} } \! . \! 
\label{eq:RxNoiseMatchMP}
\end{align}

If Tx-Rx coupling is strong, $\Z\T{}\outZ$ affects $\Z\A{}\outZ$ and $\Z\R{}\inZ$ affects $\Z\A{}\inZ$ appreciably. To our knowledge, such simultaneous matching problems have closed-form solutions only for the SISO case, e.g. see \cite{DumphartICC2017}. A possible heuristic approach is matching the Tx and Rx alternatingly with a few iterations.

An individual two-port matching network per coil is obligatory for distributed arrays but also useful to increase the matching bandwidth of co-located arrays \cite{HassanWCNC2015}. Only block-diagonal $\Z\T{}$ and $\Z\R{}$ can be realized this way and finding their PTE- or SNR-optimal values is a complicated problem for coupled arrays. 
Possible approaches include numerical optimization \cite{HassanWCNC2015} or heuristics like matching just for the diagonal of $\Z\A{}\inZ$ or $\Z\A{}\outZ$.

\subsection{MIMO Model Consistent with Physical Power}
\label{sec:PowerWaveModel}

The generators deliver currents $\i\G{} = (\Z\T{}\inZ + R \eye_{N\T{}})^{-1} \v\G{}$ and thus an average active power
$P\T{}{} = \EV{\re(\i\G{}\H \Z\inZ\T{} \i\G{})}$. We use
${\bf x} = (\re\,\Z\inZ\T{})^\f{1}{2} \i\G{}$ 
as transmit signal vector because $\EV{\|{\bf x}\|^2} = P\T{}{}$ ensures consistency with physical sum transmit power \cite[Eq.~99]{IvrlacTCS2010}. Per-generator power constraints, which occur in cooperative transmission of energy-limited sensor devices, apply to the diagonal elements $S_{n,n}$ of a matrix ${\bf S} = \re(\Z\inZ\T{}\,\EV{\i\G{} \i\G{}\H})$. Note that 
each $S_{n,n}$ depends on all generator currents because they affect the $n$-th port voltage.

As receive signal we consider the power wave ${\bf y} = \f{1}{\sqrt{R}} {\bf v}\L{}$, the natural choice for power transfer and SNR considerations. Taken all together, we obtain a MIMO model
\begin{align}
&\, {\bf y} = {\bf Hx} + {\bf n} , \\[.3mm]
& {\bf H} = \tfrac{1}{\sqrt{R}} {\bf D} (\Z\T{}\inZ + R \eye_{N\T{}}) (\re\,\Z\inZ\T{})^{-\f{1}{2}} , \\[1.1mm]
&\, {\bf n} \sim \mathcal{CN}({\bf 0}, {\bf K}) .
\label{eq:channelMatrix}
\end{align}

In the special case of weak Tx-Rx coupling and perfect power matching on both ends, the channel matrix becomes
\begin{align}
{\bf H} &= \tfrac{1}{2} \left(\re\,\Z\A{:R}\right)^{-\f{1}{2}} \Z\A{:RT} \left(\re\,\Z\A{:T}\right)^{-\f{1}{2}} .
\label{eq:Hppm}
\end{align}
When instead noise matching is used at the Rx-side, a scaling factor $\f{2R}{R + \ZOpt} (\f{\re \ZOpt}{R})^\f{1}{2}$ applies.

\section{Broadband Evaluation}\label{sec:broad}
Practical matching networks can attain the desired $\Z\T{}$ and $\Z\R{}$ values only at the circuit design frequency $\fDesign$ and deviate at $f \neq \fDesign$. This needs to be considered in broadband evaluations, for which guidelines are provided in the following.



\subsubsection{Design Phase}
Consider $f = \fDesign$ and find circuit designs for the matching networks such that $\Z\T{}(\fDesign),\Z\R{}(\fDesign)$ take values according to Sec.~\ref{subsec:match}. Appropriate lumped element designs for two-port networks include L-, T- and $\Pi$-structures; for multiport networks refer to \cite{NieTCS2014}. 

\subsubsection{Evaluation Phase}
Partition the considered $f$-range into narrow bands with center frequencies $f_k$ and width $W$ that must be chosen smaller than the channel coherence bandwidth. For each $f_k$, evaluate all $f$-dependent quantities, in particular $\Z\T{}(f_k),\Z\R{}(f_k)$ based on the fixed circuit designs. Then compute ${\bf H}_k$ from \eqref{eq:channelMatrix} and ${\bf K}_k$ from \eqref{eq:NoiseCovar} to obtain parallel MIMO channels ${\bf y}_k = {\bf H}_k {\bf x}_k + {\bf n}_k$ for the bands $k=1,\ldots,K$. Thereby, ${\bf n}_k$ and ${\bf n}_l$ are statistically independent for $k \neq l$.


\section{Biomedical Setup and Link Design}\label{sec:setup}
\begin{figure}[!ht]
  \subfloat[Setup overview]{%
    \psGD{x}{2mm}{.5mm}{$y$}    
    \psGD{y}{-.5mm}{-.5mm}{$x$} 
    \psGD{z}{-2.7mm}{0.8mm}{{$z\ [\mathrm{mm}]$}}
    \psGD{ex}{-12.2mm}{2mm}{\normalsize\textcolor[rgb]{0,0,1}{external coil array: power source \& data sink}}
    \psGD{sk}{-14.8mm}{-1.3mm}{\rotatebox{-13.9}{\normalsize\shortstack[c]{human skin\\and tissue}}}
    \psGD{sw}{-14.6mm}{-7.6mm}{\small\textcolor[rgb]{1,0,0}{\shortstack[l]{in-body\\[-.9mm]sensor nodes:\\[-.1mm]power sink \&\\[-.7mm]data source}}}
    \psGD{pr}{3.2mm}{-11.3mm}{\small\shortstack[l]{passive\\[-.7mm]relays}}
    \psGD{dl}{-8.4mm}{-1.0mm}{\small\textcolor[rgb]{.25,.25,.25}{\shortstack[r]{power\\[-.9mm]downlink}}}
    \psGD{ul}{-1.2mm}{-1.0mm}{\small\textcolor[rgb]{.25,.25,.25}{\shortstack[l]{data\\[-.3mm]uplink}}}
    \includegraphics[width=0.58\linewidth,trim=-7 0 -5 0,clip=true]{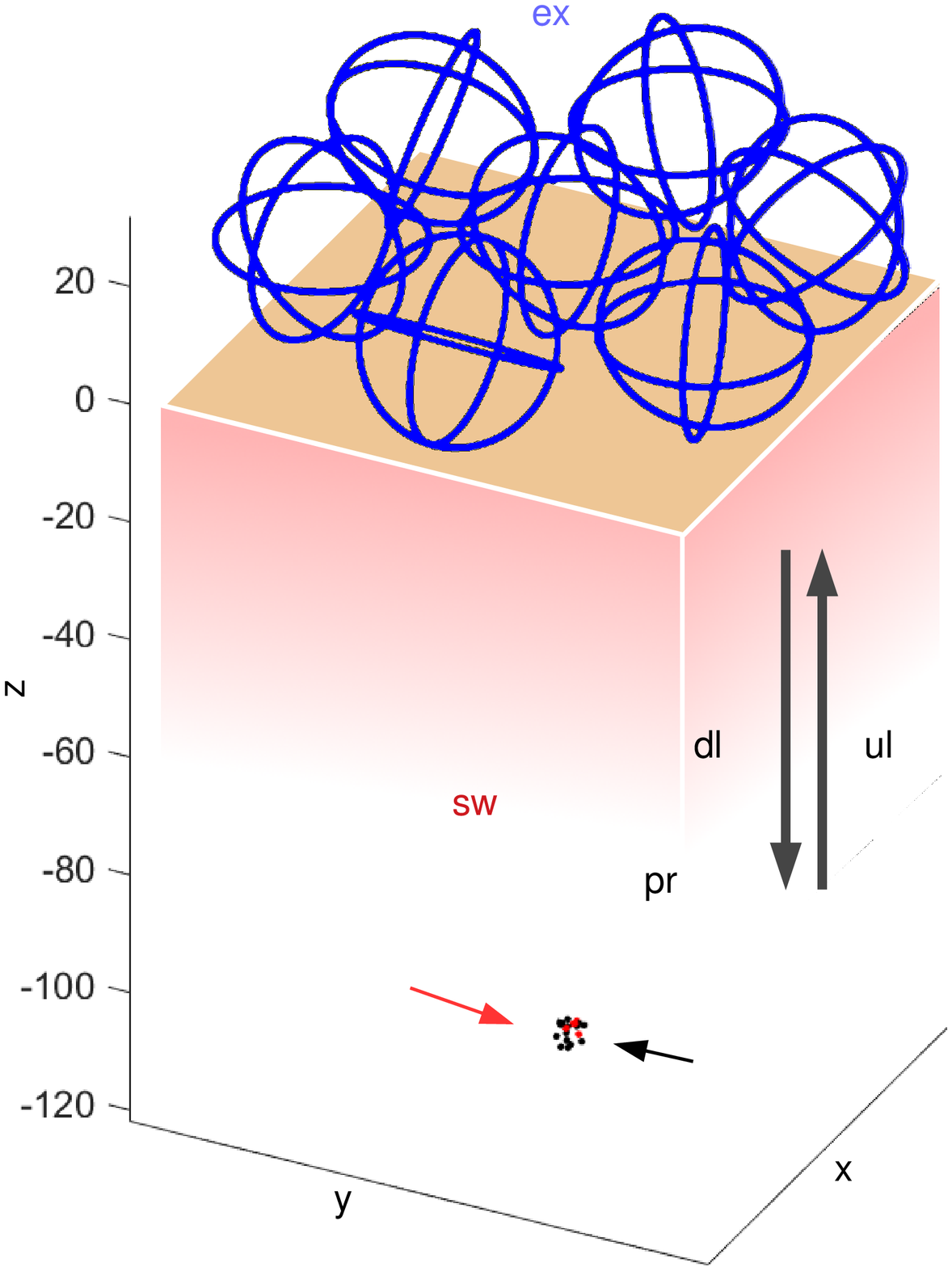}}
  \subfloat[Detail]{\begin{tabular}[b]{c}%
    \psGD{xmm}{-3.5mm}{0mm}{\scriptsize$x\ [\mathrm{mm}]$}
    \psGD{ymm}{-1.4mm}{0.3mm}{\scriptsize$y\ [\mathrm{mm}]$}
    \includegraphics[width=0.34\linewidth,trim=14.5 0 43 13,clip=true]{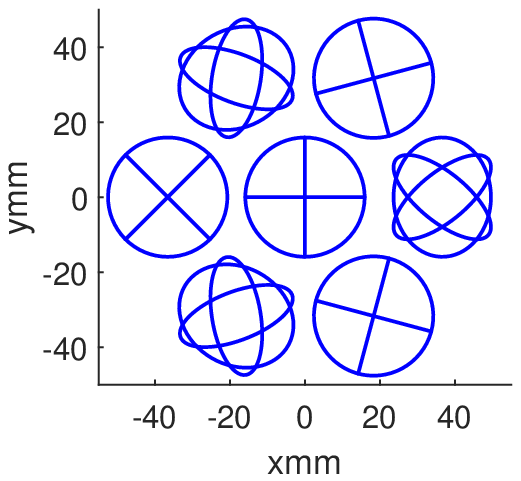} 
    \\[2mm]
    \psGD{xmm}{-3.5mm}{-.3mm}{$x\ [\mu\mathrm{m}]$}
    \psGD{ymm}{-1.8mm}{-1.7mm}{$y\ [\mu\mathrm{m}]$}
    \includegraphics[width=0.40\linewidth,trim=16 0 43 0,clip=true]{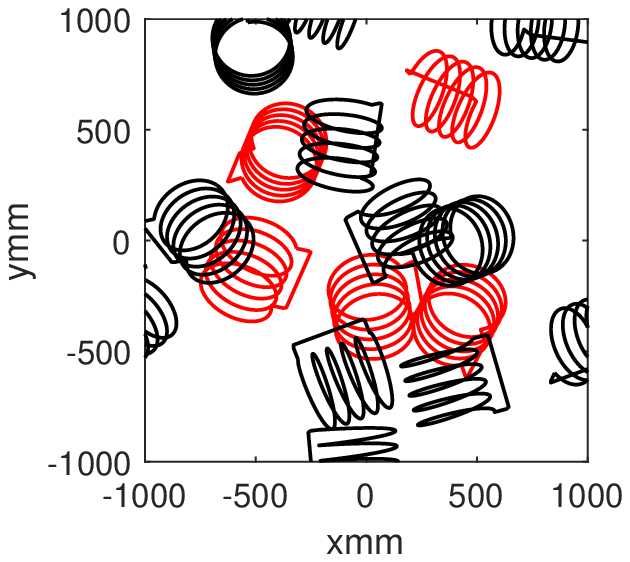} 
    \label{fig:biomedDetail}
  \end{tabular}}
\caption{Biomedical setup with an in-vivo swarm of micro-scale sensor nodes, each equipped with a multi-turn coil, located $12\unit{cm}$ beneath the skin. They receive power from and send data to an external array of 21 coils. The sensors and the accompanying resonant passive relay coils have random arrangement.}
\label{fig:biomed}
\end{figure}

As outlined, magnetic induction is a useful propagation mechanism for medical in-body devices because of its capabilities in terms of power transfer and media penetration. In the following we study the performance and feasibility limits of magneto-inductive wireless powering (downlink) and data transmission (uplink) for micro-scale devices in the exemplary biomedical in-body application illustrated in Fig.~\ref{fig:biomed}. Thereby, the micro-scale in-body sensors are located $12\unit{cm}$ beneath the skin and are each equipped with a multi-turn loop antenna (copper wire). The external device (above the skin) is both the data sink and the field source for wireless powering, for which it uses a transmit power of $1\unit{W}$. The device features a massive array of 21 coils, each with $10\unit{cm}$ circumference, in an arrangement that provides vast spatial diversity. We assume that $P_0 = 50\unit{nW}$ are required for activation of the sensor chip (cf. $450\unit{nW}$ reported in \cite{ZouJSSC2009}) and that 50\% of the excess received power is then used as transmit power in the uplink.

The system and coil parameters are chosen for best performance in this setting. The design frequency $\fDesign$ should be as large as possible for strong magnetic induction, cf. $\omega$ in \eqref{eq:MutualImpDipole}, but sufficiently low to penetrate tissue and to not be limited by $R^\text{rad}$, i.e. the wavelength should exceed the coil wire length considerably. In that regard, we choose $\fDesign = 750\unit{MHz}$ and single-turn coils with $3\unit{mm}$ thick copper wire at the external device, which lead to better results than multi-turn coils. 
The coils have a Q-factor of $266$ and are treated as electrically small (the wavelength is $40\unit{cm}$).

For the micro-scale sensors we assume single-layer solenoid coils whose height equals their diameter (henceforth called \textit{size}) with $5$ turns and a turn spacing of $1.5$ wire diameters. We calculate their resistance with \cite[Sec.~5.2.3]{Balanis2016}, inductance with \cite{MillerPIEEE1987}, and self-capacitance with \cite[Eq.~5.3]{Knight2009}.

For simplicity, we make the idealistic assumptions of full channel state information (CSI) on both ends and full-duplex operation of down- and uplink. The chosen $750\unit{MHz}$ are close to the suggestion of $\approx 1\unit{GHz}$ by \cite{PoonTAP2010} and sufficiently low-frequency to penetrate a few $\unit{cm}$ of tissue. 
Thus, we can assume free-space propagation and still obtain meaningful implications for biomedical engineering, without a detailed model for tissue and body fluids (note that $\mu_\text{r} \approx 1$ for water).

Each sensor node uses a two-port matching network of two lumped elements in L-structure. We assume that the generators of the high-complexity external device are not limited by their matching and, hence, we employ the transmit-side assumptions of \eqref{eq:Hppm} for the powering downlink. In the uplink, each coil of the external device (now receiving) has a two-port network for noise matching, implemented with a T-structure of three reactive lumped elements. The component values are optimized numerically for maximum average SNR (average in $\unit{dB}$ and over three orthogonal sensor orientations) at $\fDesign$. We refrain from a full multiport matching network as it would be ultra narrowband and vulnerable to losses and drift due to the large number of lumped elements, e.g., $N\R{} (2N\R{}+1)$ in $\Pi$-structure.
For the noise parameters we assume human body temperature for $T$ and $T_\text{A}$ (shielded environment), $\sigma_\text{iid}^2 = 0$, LNA parameters $\beta = 5\cdot 10^{-23}\unit{A}^2$ per $\mathrm{Hz}$, $R\N{} = 50\,\Omega$ and $\rho = 0.5 + 0.3j$. We use the spatial correlation model $\Phi_{mn} = J_0(kd_{mn})\,\o_m\Tr \o_n$ with Bessel function $J_0$.

\section{Numerical Results}\label{sec:effects}
\subsection{Single In-Body Sensor}%
\begin{figure}[!ht]\centering
\psGD{bo}{37.8mm}{2.2mm}{\scriptsize{\shortstack[l]{perfect matching at both ends}}}
\psGD{rs}{37.5mm}{0.6mm}{\scriptsize{\shortstack[r]{practical matching at sensor\\[-.4mm](downlink case)}}}
\psGD{rb}{20.6mm}{.7mm}{\scriptsize{\shortstack[l]{practical matching\\[-.5mm]at both ends\\[-.1mm](uplink case)}}}
\psGD{df}{-30.2mm}{-9.8mm}{\scriptsize{\shortstack[c]{design\\[-.5mm]frequency}}}
\psGD{fM}{-4.2mm}{-.1mm}{\footnotesize{$f\ [\mathrm{MHz}]$}}
\psGD{CG}{-8.0mm}{0.0mm}{\footnotesize{Channel Gain $[\mathrm{dB}]$}}
\includegraphics[width=\columnwidth,trim=13 6 25 10,clip=true]{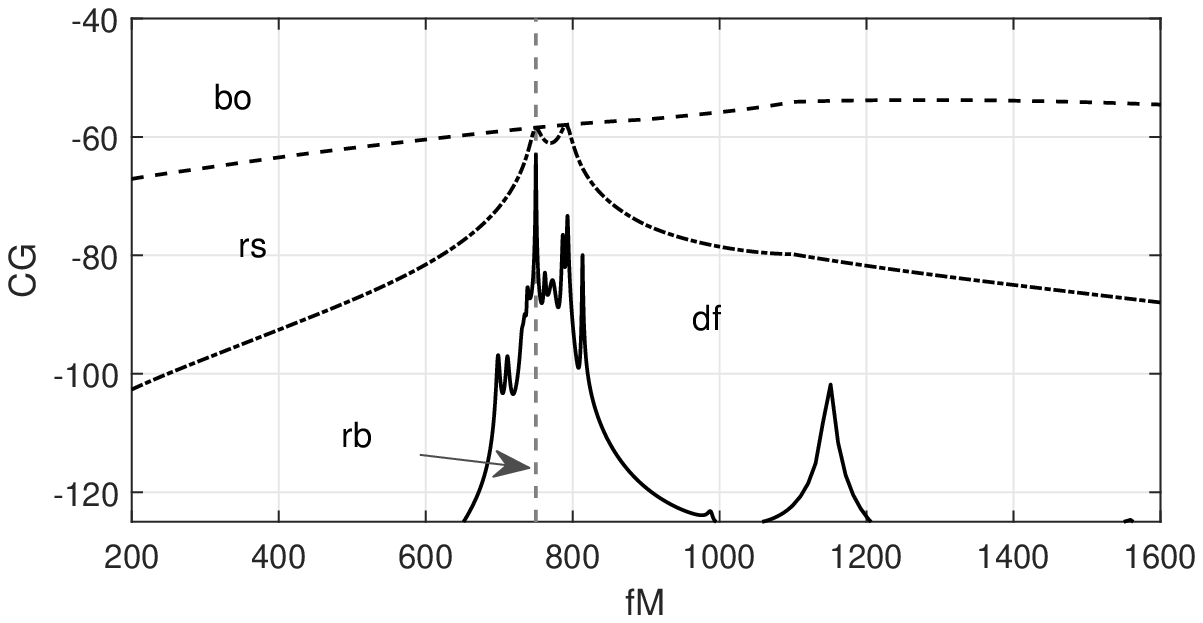}
\caption{Spectrum of the downlink channel to a micro-scale node with a $350\um$-sized coil. This addresses the channel after maximum-ratio combining at the external array. The solid line graph refers to using the two-port matching networks at the array which are assumed for the uplink. The results beyond $1\unit{GHz}$ are increasingly unreliable due to increasing electrical size.}
\label{fig:Spectrum}
\end{figure}

\label{sec:singleSens}%
%
We consider a single in-body sensor with arbitrary orientation and no passive relays.
In the downlink, the external device transmits at the design frequency and uses maximum-ratio combining for beamforming to maximize the receive power. The resultant power transfer efficiency (PTE) is $\|{\bf h}_\text{down}\|^2$ with channel vector ${\bf h}_\text{down}$. Fig.~\ref{fig:Spectrum} shows the PTE over $f$.

\begin{figure}[!t]\centering
\psGD{raa}{-4.6mm}{1.6mm}{\scriptsize{\shortstack[l]{maximum value graph\\[-.6mm]for arbitrary sensor\\[-.8mm]orientation}}}
\psGD{rai}{-3.3mm}{-0.1mm}{minimum \ldots}
\psGD{rba}{-13.4mm}{-0.55mm}{\scriptsize{\shortstack[l]{maximum value graph\\[-.6mm]for arbitrary sensor\\[-.8mm]orientation}}}
\psGD{rbi}{-6.8mm}{-1.1mm}{minimum \ldots}
\psGD{sc}{-18.0mm}{-0.9mm}{\footnotesize{Coil Size of In-Body Device $[\um\,]$}}
\psGD{dPTE}{-8.5mm}{-0mm}{\footnotesize{Downlink PTE $[\mathrm{dB}]$}}
\psGD{ulDRa}{-11.5mm}{-0mm}{\footnotesize{Uplink Data Rate $[\mathrm{bit/s}]$}}
\psGD{thr}{4.8mm}{0.3mm}{\textcolor[rgb]{1,0,0}{\scriptsize{\shortstack[r]{outage threshold with\\[-.4mm]$P_0=50\unit{nW}$\\[-.2mm]$P\T{}=1\unit{W}$}}}}
\includegraphics[width=\columnwidth,trim=21 12 31 19,clip=true]{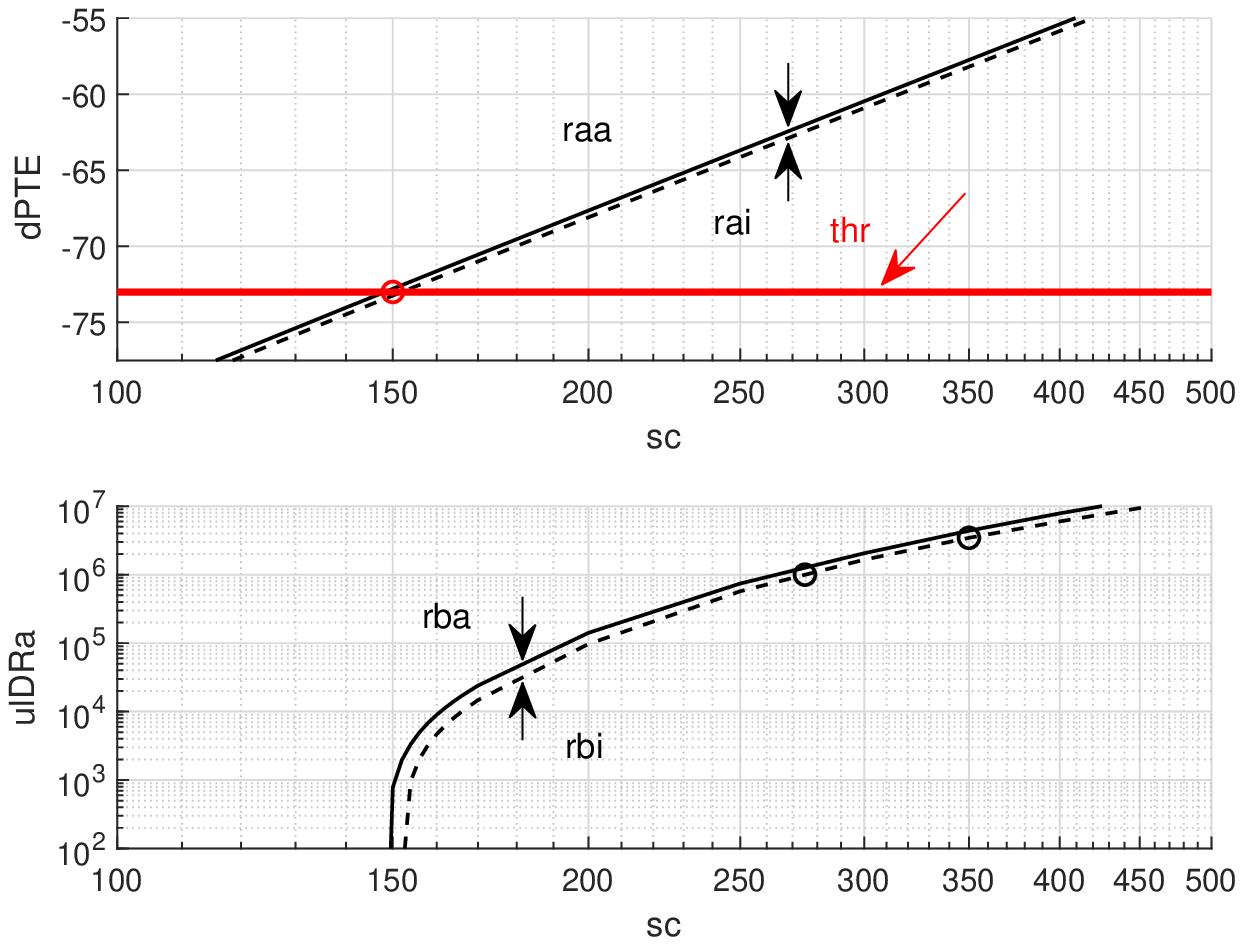}
\caption{Downlink power transfer efficiency and uplink data rate as a function of the sensor-side coil diameter (which is set equal to the coil length). The sensor is located $12\unit{cm}$ apart from the external array.}
\label{fig:SingleSensor}
\end{figure}

In the uplink, the sensor uses $P\T{,up} = \f{1}{2}(P\R{} - P_0)$ as transmit power if $P\R{} > P_0$ (otherwise, the sensor is in outage).
The now receiving array performs maximum-ratio combining in every frequency bin $k$ based on the noise-whitened channel vector ${\bf K}_k^{-1/2} {\bf h}_k$. The gains of the resultant parallel SISO channels are then used on the sensor side to find the spectral transmit power allocation $\sum_k P_k = P\T{,up}$ with the waterfilling algorithm \cite{Tse2005}. We study the uplink data rate in terms of the channel capacity
$R_\text{up} = \sum_k W \log_2(1 + P_k \cdot {\bf h}_k\H {\bf K}_k^{-1} {\bf h}_k)$.

Fig.~\ref{fig:SingleSensor} shows the evolution of uplink data rate over sensor-side coil size. We observe that a sensor can be activated and transmit data to outside the body if its coil is larger than $150\um$. At this size, $R^\text{ohm}\! = 0.52\,\Omega$, $L = 3\unit{nH}$ and $Q = 28.5$. Such size would be sufficiently small for many medical target applications \cite{Nelson2010}. With increasing size, the data rate grows rapidly as $Q$ and the the mutual impedance to the external coils increase. Coils larger than about $275\um$ promise data rates beyond $1\unit{Mbit/s}$. The sensor orientation only has a weak impact on performance due to spatial diversity and the matching design of the array. Orientation-dependent signal attenuation is mitigated even further because reactive and radiative propagation modes are phase-shifted and both significant, because $kd \approx \f{1}{3}$ in \eqref{eq:MutualImpDipole}. This is in contrary to the significant losses over unaligned near-field SISO links \cite{DumphartPIMRC2016Short}.

\subsection{Effect of Passive Relays}
\label{sec:relaying}

We investigate the effect of a randomly arranged swarm of 19 passive resonant relay coils around the sensor node. They could be placed in hopes of a performance gain or just represent nearby idle sensor nodes. Strong coupling to a relay detunes the sensor coil and shifts the resonance peaks in frequency, on the order of the sensor-coil $3\unit{dB}$ bandwidth. Likewise, dense and arbitrarily arranged swarms of passive relays cause $f$-selective channel fluctuations, i.e. fading \cite{DumphartICC2017}. We are interested in the implications for our application.

Henceforth we set the size of all sensor and relay coils to $350\um$, resulting in $R^\text{ohm}\! = 0.48\,\Omega$, $L = 7.2\unit{nH}$ and $Q = 71$. The passive relays are terminated with a capacitor for resonance at $\fDesign$ in uncoupled condition. All coil orientations are random with uniform distribution in 3D, see Fig.~\ref{fig:biomedDetail}, and the passive relay locations are sampled per coordinate from a Gaussian distribution about the sensor location (the standard deviation is 1.5 coil sizes and we resample until no coil geometries collide). We study the uplink rate with an elaborate scheme as well as a simple scheme: (i) adapt sensor matching to the coupling conditions, $f$-tuning in the downlink and waterfilling in the uplink and (ii) no matching adaptation, downlink at the design frequency $\fDesign$ and flat uplink power allocation over the $3\unit{dB}$-bandwidth of the external coil.

\begin{figure}[!b]
\vspace{-3.5mm}
\subfloat[One sensor]{
  \centering
  \psGD{ECDF}{-5.0mm}{1.0mm}{\footnotesize{Empirical CDF}}
  \psGD{ulDRa}{-11.5mm}{-.2mm}{\footnotesize{Uplink Data Rate $[\mathrm{Mbit/s}]$}}
  \psGD{sm}{0.12mm}{-0.1mm}{\color[rgb]{0.85,0.33,0.10}{static matching,}}
  \psGD{ss}{0.12mm}{0.1mm}{\color[rgb]{0.85,0.33,0.10}{simple signaling}}
  \psGD{rs}{0.0mm}{0.0mm}{re-matched sensor,}
  \psGD{sa}{0.0mm}{0.3mm}{spectrally aware signaling}
  \includegraphics[width=\columnwidth,trim=0 0 0 7,clip=true]{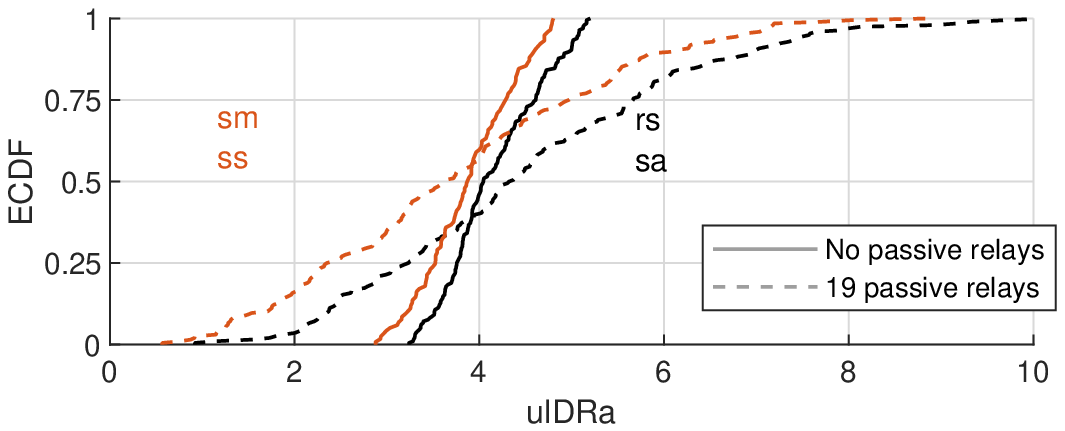}
  \label{fig:RateCDFsSingle}}
\ \\
\subfloat[Five sensors, cooperative transmission]{
  \centering
  \psGD{ECDF}{-5.0mm}{1.0mm}{\footnotesize{Empirical CDF}}
  \psGD{ulDRa}{-11.5mm}{-.2mm}{\footnotesize{Uplink Data Rate $[\mathrm{Mbit/s}]$}}
  \psGD{sm}{0.25mm}{-0.3mm}{\color[rgb]{0.85,0.33,0.10}{static matching,}}
  \psGD{ss}{0.25mm}{-0.1mm}{\color[rgb]{0.85,0.33,0.10}{simple signaling}}
  \psGD{rs}{0.0mm}{0.0mm}{re-matched sensors,}
  \psGD{sa}{-0.85mm}{0.3mm}{spectrally aware signaling}
  \includegraphics[width=\columnwidth,trim=0 0 0 7,clip=true]{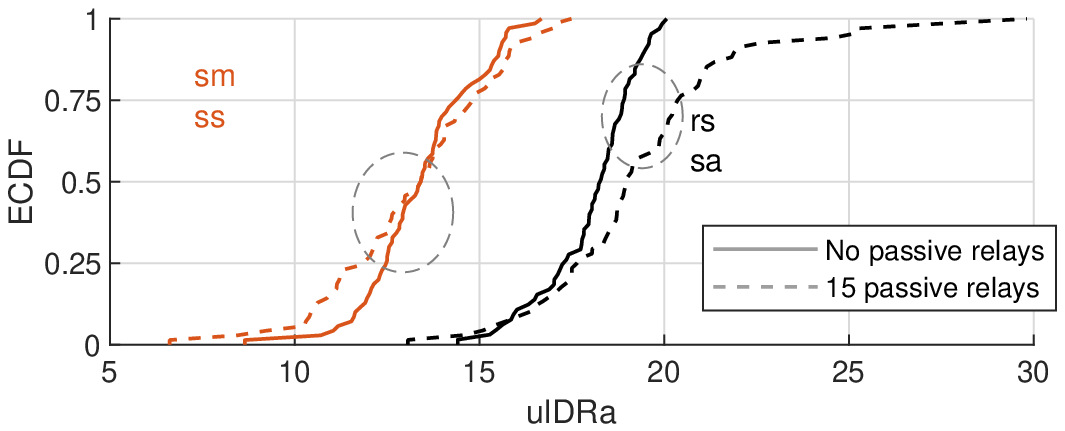}
  \label{fig:RateCDFsCoop}}
\caption{Uplink data rates from one or multiple (cooperating) in-body sensors, with and without nearby passive resonant relay coils in random arrangement, to an external device at $12\unit{cm}$ distance. Either case considers 20 in-body coils of $350\um$ size. The external device uses $1\unit{W}$ to supply power wirelessly. The results are shown as cumulative distribution function (CDF).}
\label{fig:RateCDFs}
\end{figure}

Fig.~\ref{fig:RateCDFsSingle} shows the resulting uplink rates for many realizations of the random swarm geometry. We observe that the relays indeed cause a fading effect. Clearly, meeting the \mbox{$f$-dependent} fluctuations and detuning with appropriate measures improves the uplink;
exploiting selective fading via transmit CSI is well-established for radio links \cite{Tse2005}.
Still, the presence of relays is detrimental in about 40\% of cases.

\subsection{Cooperative Transmission with Other Sensors}

Consider that the sensor node of interest is accompanied by other sensor nodes and possibly passive relays, and all activated sensors cooperate in the uplink. We assume that the sensors can establish phase synchronization because of the sub-$\mathrm{GHz}$ operation. Furthermore, due to the strong links between neighboring sensors, we assume that their data exchange rates do not pose a bottleneck and thus the data rate of the cooperative uplink is determined by the final-hop rate. We assume the presence of 15 passive relays. All relay and sensor locations and orientations are sampled as described in Sec.~\ref{sec:relaying} and an example arrangement is shown in Fig.~\ref{fig:biomedDetail}.

In the downlink, we use maximum-ratio transmit combining at the external array to maximize the received sum power $\|{\bf y}\|^2$ over this MIMO channel, although subject to $|y_1|^2 \geq P_0$ to activate the sensor of interest whenever possible. This way, vast power is fed to sensors with a good channel, with the idea that those sensors should see a good channel in the uplink, which depends on array matching and noise statistics though.

Again, we compare a simple scheme with no matching adaptation, downlink operation at $\fDesign$ and simplistic uplink power allocation (as in Sec.~\ref{sec:relaying}) to a more sophisticated scheme which adapts the matching of the sensors to the coupling conditions and the signaling to the frequency-selective channel. For the latter, spectral power allocation in the uplink is concerned with $\sum_k P_{k,n} = P_{\text{T,up},n}$ per sensor $n = 1 \ldots 5$. Since this case has no known analytical solution for the rate-optimal allocation \cite{GoguriICASSP2016} we choose a heuristic approach: each sensor $n$ allocates $P_{k,n}$ via waterfilling over the hypothetical parallel SISO channels that would arise when all other sensors are silent (but present) and the receive combining is as in Sec.~\ref{sec:singleSens}. We obtain an achievable uplink rate of $R_\text{up} = \sum_k W \bar{R}_k$ where $\bar{R}_k = \max_{\bf Q} \log_2 \det(\eye_{N\R{}} + \bar{\bf H}_k {\bf Q} \bar{\bf H}_k\H)$  is found by solving the optimization problem subject to ${\bf Q} \succeq {\bf 0}$ and the per-node power constraints given by the $P_{k,n}$, which are incorporated according to Sec.~\ref{sec:PowerWaveModel}. Thereby, $\bar{\bf H}_k = {\bf K}_k^{-1/2} {\bf H}_k$ is the noise-whitened channel matrix in frequency bin $k$.

Fig.~\ref{fig:RateCDFsCoop} shows that the considered cooperative scheme yields a three- to fourfold increase of uplink data rate over the non-cooperative case in Fig.~\ref{fig:RateCDFsSingle}, with the same amount of source power. Node cooperation can exploit the location-dependency of swarm-induced signal fluctuations, which usually improve the channel only for a subset of the nodes. Furthermore, the difference between the simplistic and elaborate approach is now more pronounced: in about 93\% of cases, the elaborate scheme manages to utilize the passive relaying effect for a (sometimes significant) performance gain by exploiting the spectral and spatial channel variations.

\section{Summary}\label{sec:sum}
We presented a general system model for magneto-inductive networks in order to study the performance limits of medical microsensors which receive power and transmit data via magnetic induction. For $12\unit{cm}$ distance to the power source and data sink, the determined minimum coil size (copper wire) is about $150\um$, while $275\um$ already allow for $1\unit{Mbit/s}$ uplink data rate. We furthermore showed that the passive relaying effect in dense swarms of resonant nodes can be utilized for performance gains with adaptive matching, spectral power allocation, and/or node cooperation. Future research should extend the study to health regulations and propagation effects in tissue, absence of CSI, advanced materials (e.g. graphene) and compare to acoustic and optical approaches.

\appendices

\newpage
\section*{Appendix: In- and Output Impedance Matrices}
\label{app:PortImp}
When all inputs are terminated as shown in Fig.~\ref{fig:CircSignal}, the impedance matrices at the output ports are
\begin{align}
\Z\T{}\outZ &= \Z\T{:A} - \Z\T{:AG} \left( \Z\T{:G} + R\eye_{N\T{}} \right)^{-1} \Z\T{:AG}\Tr \\
\Z\A{}\outZ &= \Z\A{:R} - \Z\A{:RT} \left( \Z\A{:T} + \Z\T{}\outZ  \right)^{-1} \Z\A{:TR} \\
\Z\R{}\outZ &= \Z\R{:L} - \Z\R{:LA} \left( \Z\R{:A} + \Z\A{}\outZ  \right)^{-1}  \Z\R{:LA}\Tr .
\end{align}
With terminated outputs, the input impedance matrices are
\begin{align}
\Z\T{}\inZ &= \Z\T{:G} - \Z\T{:AG}\Tr \left( \Z\T{:A} + \Z\A{}\inZ \right)^{-1} \Z\T{:AG} \\
\Z\A{}\inZ &= \Z\A{:T}  - \Z\A{:TR}  \left( \Z\A{:R}  + \Z\R{}\inZ \right)^{-1} \Z\A{:RT} \\
\Z\R{}\inZ &= \Z\R{:A} - \Z\R{:LA}\Tr \left( \Z\R{:L} + R \eye_{N\R{}} \right)^{-1}  \Z\R{:LA} .
\end{align}



\IEEEtriggeratref{0}
\bibliographystyle{IEEEtran}
\bibliography{ms}


\end{document}